\documentclass[aps,prl,twocolumn,10 pt,superscriptaddress,showpacs]{revtex4}
\usepackage{amssymb}
\usepackage{graphicx}

\begin{document}

\title{Dominant role of impurity scattering over crystalline anisotropy for magnetotransport
 properties in the quasi-1D Hollandite $Ba_{1.2}Rh_{8}O_{16}$.}
\author{Alain Pautrat}
\affiliation{Laboratoire CRISMAT, UMR 6508 du CNRS, ENSICAEN et Universit\'{e} de Caen, 6 Bd Mar\'{e}chal Juin, F-14050 Caen 4.}
\author{Wataru Kobayashi}
\affiliation{Waseda Institute for Advanced Study, Waseda University, Tokyo 169-8050, Japan.}

\begin{abstract}

Angular magnetotransport measurements have been performed to tackle the origin of 
the magnetoresistance in the quasi-1D Hollandite $Ba_{1.2}Rh_{8}O_{16}$.
Three samples of different impurities amount were measured. We observe that the low temperature resistivity upturn is not
 due to a charge density wave transition,
 and a dominant role of impurities scattering for low temperature transport properties is instead demonstrated.
The components of magnetoresistance were separated by using the Kohler plot and the angular dependency of the
 resistance under magnetic field. It shows the major contribution of an isotropic,
 likely spin driven, negative magnetoresistance. Galvanomagnetic characteristics are then consistent with
 a Kondo effect and appear to be essentially 3D at low temperature.

\end{abstract}

\newpage
\maketitle

\section{Introduction}
 
$Ba_{1.2}Rh_{8}O_{16}$ is a quasi-one dimensional Hollandite \cite{wata1}. In this oxide, Rhodium presents a mixed valency
with $Rh^{3+}$ and $Rh^{4+}$.
Interestingly for a low dimensional oxide, it has a pronounced metallic character along the chains direction, with a room temperature
 resistivity of about 1 $m\Omega.cm$.
Phonon scattering seems to take a major part in the dissipation, as the temperature variation follows that deduced from Bloch-Gruneisen
functions with a reasonable Debye temperature \cite{wata1}. Note that if this result is expected in the case of a simple 3D metal,
 it is $\textit{a priori}$
more surprising for a quasi-1D system where electron electron umklapp scattering can take a large part in the dissipative processes \cite{1Dresis}
 and where corrections due to the non spherical Fermi surface can be important.
 In addition, a small upturn of the resistivity is observed at low temperature.
 It is apparently associated with a small negative magnetoresistance, whose origin remains unexplained \cite{wata1}.

Electronic instabilities are expected in low dimensional conductors. The first one
 is superconductivity whose manifestation (the resistance R=0 and the Meissner effect) is unambiguous and is clearly not
 observed in $Ba_{1.2}Rh_{8}O_{16}$ at least for $T>1.8$ K.
 Considering the extreme low residual resistivity needed to observe superconductivity
 in non conventional oxide superconductors such as $Sr_2RuO_4$ \cite{SrRuO}, the level of disorder in $Ba_{1.2}Rh_{8}O_{16}$
 is likely in the range where pair breaking is expected.
  More directly related to the low dimensionality are the Nesting properties of the Fermi surface which  
 lead to a Peierls type transition and to a charge-density wave. A direct consequence is a metal to insulator transition in the pure 1D case
 or eventually a more complex transition
 due to incomplete nesting in higher dimensions. The upturn of resistivity of $Ba_{1.2}Rh_{8}O_{16}$ at low temperature could
 be a manifestation of such a charge density wave instability due to the low electronic dimensionality. In this case,
 non linear transport properties characteristics of the CDW
 pinning and sliding are expected \cite{gruner}.
Other possibilities for explaining the resistivity upturn (and the weak negative magnetoresistance) are Kondo effect \cite{kondo} and weak localization phenomena \cite{weak}. 
In general, several complementary measurements are needed because it is difficult to disentangle these different magnetoresistive processes.
To go further into the magnetotransport properties of this interesting compound, 
we have measured in details three samples of $Ba_{1.2}Rh_{8}O_{16}$ of the same
 batch but presenting different residual resistivities, \textit{i.e.} different amount of impurities.

\section{Experimental}
 
Single crystals of $Ba_{1.2}Rh_{8}O_{16}$ were grown using a flux method as detailed in \cite{wata1}.
Unit cell parameters are a= 10.446 , b= 3.051 and c= 9.424 $\AA$ in the monoclinic cell.
Crystals are needle-like with the longest dimension along the b-axis, and have typical dimensions of 400$\times$25$\times$25 $\mu m^3$ (Fig.1).
The precise dimensions of the measured crystals are given in the table I.

\begin{table}[h]
\begin{center}
\begin{tabular}{|c|c|c|c|}
\hline
Sample & $Section (\mu m)^2$ & $d_{contacts} (\mu m)$ & $RRR$ \\
\hline
$\sharp 1$   & 751 & 196 & 8.4 \\
\hline
$\sharp 2$  & 942 & 259 & 11.1 \\
\hline
$\sharp 3$ & 682 & 132 & 13.3 \\
\hline
\end{tabular}
\caption{Samples measured, typical sizes and residual resistivity ratio $RRR=\rho (300K)/\rho (T_{min})$}
\end{center}
\vspace{-0.6cm}
\end{table}

Four gold wires of diameter 20 $\mu m$ were attached to the sample with silver paste to perform four probes measurements.
 To allow for a small diffusion of the silver and a good contact resistance, the sample was heated at 673 K during 10 min.
 Magnetotransport measurements were performed in a physical properties measurements system (PPMS, Quantum Design), equipped with a 14T superconducting solenoid and a vertical
 rotator for sensitive angular measurements. For noise and non linear transport measurements used to check the CDW properties, a home made set-up was used \cite{Jo}.  
 
\section{Low temperature transport properties}
As shown in Fig.2 and as previously reported \cite{wata1}, $Ba_{1.2}Rh_{8}O_{16}$ exhibits a metallic resistivity from $T=400 K$ to a low temperature $T_{min}$.
At the lower temperatures $T<T_{min}$, a small upturn of the resistivity can be observed (see the inset of Fig.2). Its origin is unclear,
 but a low dimensional instability such as a CDW can provide an attractive interpretation \cite{wata1}.
 We have measured three samples which present different residual resistivity ratio (RRR)
corresponding to different strength of impurity scattering for each. In order to provide a qualitative estimation of the electronic mean
 free path $\ell$, we use a simple Drude approach which gives $\ell_{300K}\sim h/(b.N.e^2.\rho_{300K})$,
with $h$ the Planck constant, $e$ the elementary charge, $N\sim 1.02$ $10^{22} cm^{-3}$ the carriers density deduced
 from Hall effect measurements \cite{wata1}, and $b$ the b-axis parameter. Assuming that impurities are diluted enough to use the Matthiessen rule, we obtain
 $\ell_{T_{min}}=(RRR-1)\ell_{300K}$.
The cleanest sample, the sample $\sharp3$, presents a value $\ell(T_{min})\sim 11$ nm. Taking into account the quasi-1D character of the sample,
 a more specific approach from the Boltzmann equation would yield 
$\ell_{300K}\sim h.a.c/(4.e^2.\rho_{300K})$ where $a$ and $c$ are a and c-axis parameters respectively \cite{horii}.
 We find $\ell(T_{min})\sim 7.5$ nm for the sample $\sharp3$.
 The values of the apparent mean free path are of the same order of magnitude with the two approaches. The value obtained with the Drude 
formula will be used for analyzing the qualitative variation of $\ell$ between the three samples.


Disorder is in general not favorable to the condensation of a CDW, and its transition temperature is expected to decrease 
when the mean free path decreases \cite{CDW}.
 It can be observed in Fig.3 that $T_{min}$ is contrarily decreasing when the mean free path increases, meaning that the disorder favors the transition.
 In addition, we do not observe any characteristics
 of the transport properties in a CDW state (non linear transport is not observed as far as the sample is in thermal equilibrium
 and voltage noise is always measured
 resolution limited ($< 0.5 nV/\sqrt{Hz}$)).
 We conclude that the low temperature upturn of the resistivity
 is not due to CDW transition. As shown in the inset of fig.2, the thermal variation of the resistivity is $\Delta\rho(T)\sim - A \times logT$.
 This is observed in Kondo metals at low temperature for $T>>T_K$, $T_K$ being the Kondo temperature.
 To the extent that the mean free path at low temperature is inversely proportional to the impurity concentration
 as in simple metals (with dilute impurity concentration), one finds that the slope $A$ increases linearly with the impurity concentration (Fig.3),
 as also expected for Kondo resistivity in the
 high temperature limit.


\section{Low temperature magnetotransport properties}

The $Ba_{1.2}Rh_{8}O_{16}$ exhibits apparently two types of magnetoresistance (MR).
 The first one is positive and is observed for $T<T\sim20K$ down to low temperature. 
Classically, the galvanomagnetic properties of a metal are contained in the product $\omega\tau$,
 where $\omega$ is the cyclotron frequency of charge carriers across the Fermi surface and $\tau$ 
is the mean free time proportional to the mean free path.
  In this conventional picture, the MR is higher when the mean free path increases. Comparing our
 three samples, it can be observed in the Fig.4 that the positive MR is indeed increasing with the mean free path.
 The positive MR that we measure is quadratic with the magnetic field for moderate applied field values,
 as expected for a Lorentz-Force driven MR. It is linear for higher fields and non saturating (Fig.5) for reasons that deserves more attention.
 A consequence of a simple galvanomagnetic coupling (with one relaxation time) is that
 the MR curves measured at different temperatures makes a single curve when plotting
 $\Delta R/R_0=f(\omega\tau)=f(B/\rho_0)$.
 This is known as the Kohler rule \cite{kohler}. We observe here that this Kohler rule is fulfilled
 for $T>T_{min}$ but not for $T<T_{min}$ (Fig.6) showing that another magnetoresistive process is appearing. 
  
The low temperature upturn of resistivity is indeed associated with a negative MR which grows when decreasing the temperature.
 This negative MR is increasing when the RRR decreases, what is the opposite of a mean free path effect
 and clearly differentiates the negative MR from the normal (Lorentz force driven) MR.
  It implies also that the negative MR is driven by impurities scattering. 
 The pure negative MR has been extracted after correcting the
total MR by the normal MR
 which has been extrapolated for $T<T_{min}$ using the Kohler rule.
For the Kondo effect, the MR is predicted to vary approximately as $<S>^2$, the square of thermal average of localized moment,
 for temperatures $T>T_K$ \cite{giordano, monot} what is the relevant case here. The field dependence of the negative MR should be very
 close to the square of the Brillouin function, as we observe in Fig.7. The temperature dependence of the negative MR is also well explained
 by the factor (B/T) contained in the Brillouin function.
 
\section{Angular dependence of the resistance}
 
 The resistivity has been measured as function of the angle $\theta$ between the magnetic field $B$ and the transport current $I$,
 which was applied along the longest dimension of the crystal (b-axis). 
 In our convention, $\theta=0$ for $I$ perpendicular to $B$. The angular dependence of the resistivity presents here a systematic twofold symmetry.
 For $T>T_{min}$ but also as soon as the positive MR dominates over the negative MR, the resistivity
is maximum for $\theta=0$, as expected for the ordinary MR (OMR) due to Lorentz force (see Fig.8 for $B>10T$). 

 When the negative MR dominates, an opposite angular dependence develops itself, recalling the anisotropic MR (AMR) (see Fig.8 for $B<10T$).
The origin of the AMR is in the spin-orbit coupling,
 and it leads to a minimum (maximum) of resistance when the transport current is perpendicular (parallel) to the magnetization.
 The angular dependence of the AMR is in the simplest form given by \cite{AMR}:
 
 \begin{equation}
 \rho=\rho_0+\Delta\rho(1-cos^2\theta_M)
 \end{equation}
 
 $\rho_0$ is the isotropic resistivity, $\theta_M$ is the angle between current and magnetization and $\Delta\rho$ is the difference between 
the two extrema of the resistivity.
 As shown in Fig.9 for a particular field value, we obtain systematic good fits using the equation (2) with $\theta=\theta_M$. It implies that the magnetization 
follows the direction of the applied magnetic field, and that the crystalline and form
 anisotropies are easily overcome. 
At first sight, it is surprising considering that the samples are needle-like and present a large form anisotropy between the parallel and
 perpendicular geometries
(we estimate a ratio of demagnetization coefficient of about 50 from geometrical consideration). This is however consistent
 with the absence of magnetic ordering and 
with a low paramagnetic susceptibility as usually observed in Rhodium oxides \cite{rhodium, notebene}. Note that $Rh^{3+}$ is expected to be low spin,
 non magnetic ($S=0$). It should break magnetic interactions in the chains, allowing for only small magnetic units and resulting in a moderate AMR.
This AMR indeed represents only one tenth of the total negative MR. As a main point, $R(B)$ for $\theta=0$ deg and
 $\theta=90$ deg at $2K$ have similar
 variation and order of magnitude.
We conclude that the anisotropic parts of the MR are of the order of 10 \% and arise from AMR or OMR contribution (Fig.8).
The main contribution is clearly isotropic, and is associated with spin disorder.
 
Finally, we find that despite the quasi 1D structure of $Ba_{1.2}Rh_{8}O_{16}$, definitive signs of 1D electronic phenomena
 are not observed at low temperature.
Hollandites are structurally close to 1D because $BO_6$ octahedra ($B=Rh$ here) are joined by corner along the chains direction and
 by edges along the other direction \cite{hollandite}. The interchain coupling is then moderate but still exists. A non negligible interchain hopping can be anticipated.
 In such a case, the Fermi surface is not consisting in two parallel planes as in the pure 1D case, but these latter are warped
 and allow for some single-particle hopping. It leads 
to electronic properties of higher dimension than 1D at low temperature,
 specially if electronic correlations are not too strong  \cite{1D, giam}.
 In particular, 1D to 3D crossover can be expected when the thermal energy decreases.
 As shown in the inset of Fig.10, a notable anisotropy is observed between the transverse and the longitudinal resistivity,
 with a maximum which emerges at an intermediate temperature.
 In particular, the transverse resistivity,\textit{i.e} perpendicular to the needle direction, varies in a way consistent with a 3D-1D crossover.
 The high temperature of the transverse resistivity is characterized by $d\rho/dT<0$ (Fig. 10).
 It is compatible with an incoherent interchain hopping as observed in the
 Bechgaard salts \cite{Beechgard}. The low temperature transverse resistivity
 is metallic, as expected if interchain coupling takes place when thermal energy
 becomes lower than the warping of the Fermi surface.
A similar anisotropy in the carrier transport properties was reported in $BaRu_6O_{12}$ \cite{Baru}
 ($\rho_{ac}/\rho_{b}\sim$ 10), another Hollandite with a quasi-1D structure. $BaRu_6O_{12}$ was proposed to be a quasi 1D electronic oxide close to a
 quantum phase transition between two different ground states, metallic and weakly localized. The major argument was the extreme sensitivity
 of transport properties to disorder at low temperature. $BaRu_6O_{12}$ presents a positive MR at low temperature, which magnitude increases with the RRR,
 resulting indeed in a tendency to stabilize weak localized state under magnetic field.
 We have shown that the changes in the positive MR are due to a mean free path effect in $Ba_{1.2}Rh_{8}O_{16}$, with an angular dependence consistent
 with a normal, Lorentz force driven, MR. It suggests that the sensitivity of galvanomagnetic properties to disorder at low temperature in
 our quasi-1D Hollandite has no link with a quantum phase transition.
There is an additional contribution in $Ba_{1.2}Rh_{8}O_{16}$: the negative MR associated with the Kondo-like resistivity upturn
 which dominates at low temperature.
 It tends to show that the impurity
 scattering is magnetic in our sample, and non magnetic in $BaRu_6O_{12}$.
 Reasons for this difference are for the moment unclear, specially because a too small
 amount of magnetic impurities can be hard to detect. A specific feature of $Ba_{1.2}Rh_{8}O_{16}$ is the incommensurate modulation
 of Barium position \cite{wata1}. The role of local fluctuations in charge valence induced by the incommensurability
 could be an interesting track but has to be clarified with a systematic comparison of transport properties of
 isostructural commensurate and incommensurate compounds.
 
\section{Conclusion}
In summary, we have studied the magnetotransport properties of hollandite $Ba_{1.2}Rh_{8}O_{16}$.
 An isotropic component with the characteristics of a Kondo effect is observed at low temperatures. 
It appears in addition to a classical galvanomagnetic component. Both are sensitive to disorder, but only the latter shows the evolution expected for a mean free path effect.
Examining the different measurements, we propose that impurity scattering of magnetic origin plays a major role compared to others Hollandite,
 and that the low temperature transport properties have here no relation to a quantum critical point.
 At first sight, the isotropic form of magnetotransport characteristics seems to
contradict the 1D structural character of the sample, but is however not inconsistent with the expected behavior of a quasi-1D conductor.
The electronic anisotropy of $Ba_{1.2}Rh_{8}O_{16}$, if notable, seems to low to preserve pure 1D properties at low temperatures.
 The genuine origin of the magnetic scattering term is unknown, and a possible role of the structural incommensurate modulation has to be studied.   
 Further work will be required to understand this peculiar point.

Acknowlegments: A.P thanks Olivier Perez (CRISMAT) for helpful comments on the incommensurate structure of $Ba_{1.2}Rh_{8}O_{16}$.

\newpage
\begin{figure}[t!]
\begin{center}
\includegraphics*[width=8.0cm]{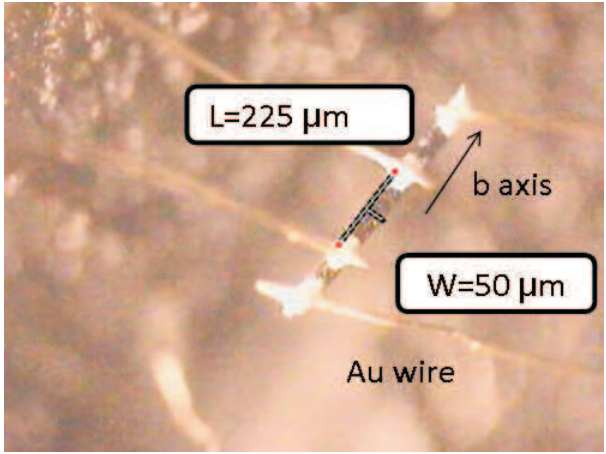}
\end{center}
\caption{(Color Online) Picture of the 1D Hollandite $Ba_{1.2}Rh_{8}O_{16}$ with the contacts. }
\label{fig.1}
\end{figure}

\begin{figure}[t!]
\begin{center}
\includegraphics*[width=8.0cm]{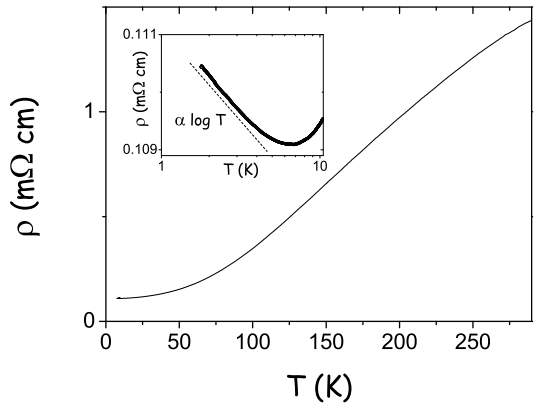}
\end{center}
\caption{Linear resistivity of $Ba_{1.2}Rh_{8}O_{16}$ as function of the temperature (sample $\sharp3$). In the inset is shown the linear-log 
of the resistivity versus temperature at low temperature. }
\label{fig.2}
\end{figure}

\begin{figure}[t!]
\begin{center}
\includegraphics*[width=8.0cm]{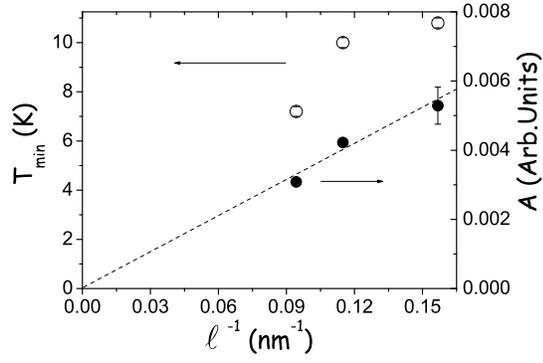}
\end{center}
\caption{$T_{min}$ and the slope $A$ (of $\rho$ vs $logT$) as function of $\ell^{-1}$. It scales at first order as the impurities concentration. The dashed line indicates a linear variation of $A$ versus $\ell^{-1}$. }
\label{fig.3}
\end{figure}

\begin{figure}[t!]
\begin{center}
\includegraphics*[width=8.0cm]{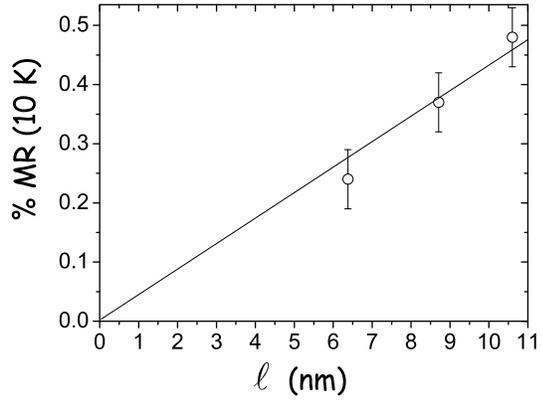}
\end{center}
\caption{Positive MR at $T=10K$, $B=7T$, as function of the electronic mean free path.}
\label{fig.4}
\end{figure}

\begin{figure}[t!]
\begin{center}
\includegraphics*[width=8.0cm]{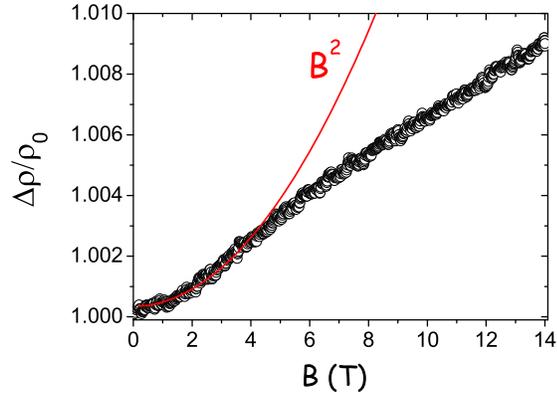}
\end{center}
\caption{(Color Online) Normalized resistivity $\Delta \rho/\rho_0$ as function of the magnetic field at $T=10K$ for the sample $\sharp 3$, the line is the $B^2$ variation. }
\label{fig.5}
\end{figure}

\begin{figure}[t!]
\begin{center}
\includegraphics*[width=8.0cm]{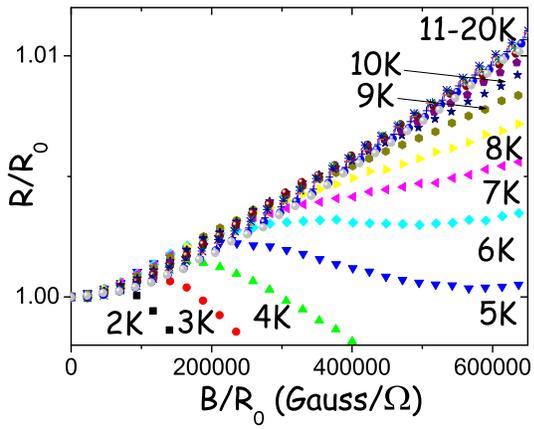}
\end{center}
\caption{(Color Online) Kohler plot $R/R_0=f(B/R_0)$ for different temperatures between $20K$ and $2K$ ($B\perp I$).}
\label{fig.6}
\end{figure}

\begin{figure}[t!]
\begin{center}
\includegraphics*[width=8.0cm]{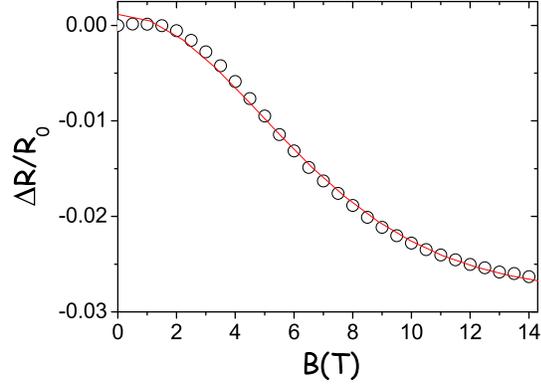}
\end{center}
\caption{(Color Online) Negative magnetoresistance $\Delta  R/R_0$ at $T=2K$ (corrected from the positive classical MR)($B\perp I$).
 The line is the functional form of the square of the Brillouin function,
 as expected for Kondo scattering in the high temperature limit.}
\label{fig.7}
\end{figure}

\begin{figure}[t!]
\begin{center}
\includegraphics*[width=8.0cm]{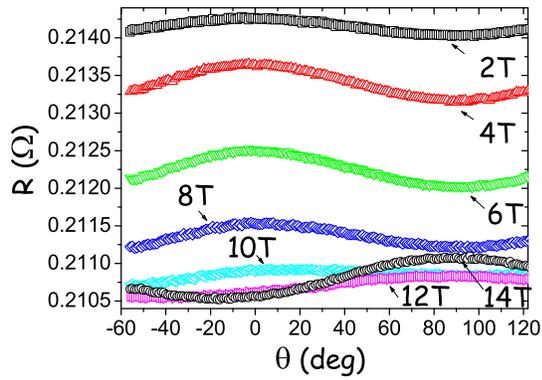}
\end{center}
\caption{(Color Online) Angular dependence of the resistance for field ranging from $2T$ to $14T$ at $T=2K$.
 Note the inversion of the angular dependence for $B\sim 10T$.}
\label{fig.8}
\end{figure}

\begin{figure}[t!]
\begin{center}
\includegraphics*[width=8.0cm]{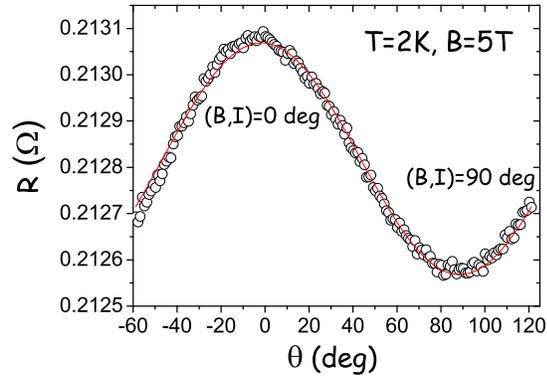}
\end{center}
\caption{(Color Online) Angular dependence of the resistance in the negative MR regime ($B=5T$, $T=2K$). The solid line is a fit using equation (2) 
with $\delta R/R_0=0.0023$.}
\label{fig.9}
\end{figure}

\begin{figure}[t!]
\begin{center}
\includegraphics*[width=8.0cm]{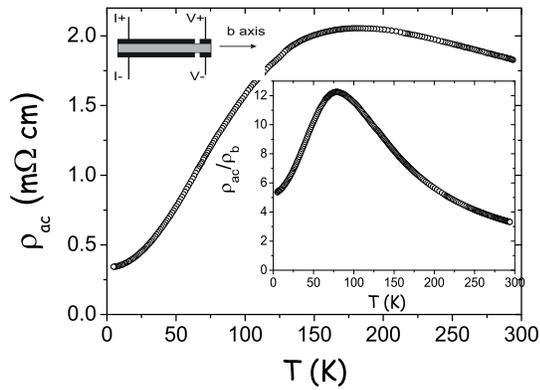}
\end{center}
\caption{Transverse resistivity $\rho_{ac}$ (perpendicular to the needle direction) as function of the temperature. Inset: Temperature dependence of the ratio between transverse and longitudinal resistivities.
Also shown is a schematic view of the contacts geometry used for the measurements of the transverse resistivity.}
\label{fig.10}
\end{figure}

\begin{references}
\bibitem{notebene}The sample are extremely small needles of typical diameter  20-40 $\mu m$ and length 400 $\mu m$, and macroscopic
 magnetic measurements yield no signal due to the lack of volume and expected small susceptibility values.
\end{references}
\end{document}